\documentclass[useAMS,usenatbib,times,letter,amssymb]{mn2e}
\usepackage{epsfig,times,amssymb}

\title[The virial to stellar mass ratio in massive galaxies]{A weak lensing estimate from GEMS of the virial to stellar mass ratio in massive galaxies to $z \sim 0.8$ }
\author[Heymans et al.]{Catherine
  Heymans$^{1,2}$\thanks{heymans@physics.ubc.ca}, 
Eric F. Bell$^2$, Hans-Walter Rix$^2$, 
Marco Barden$^2$,
Andrea Borch$^{2,3}$, \newauthor
John A.R. Caldwell$^{4}$, Daniel H. McIntosh$^5$, 
Klaus Meisenheimer$^2$, 
Chien Y. Peng$^6$, \newauthor
Christian Wolf$^{7}$, Steven V. W. Beckwith$^{6,8}$,
Boris H\"au\ss ler$^2$,  
Knud Jahnke$^2$, \newauthor
Shardha Jogee$^9$,
Sebastian F. S\'anchez$^{10}$, Rachel Somerville$^2$ \&  
Lutz Wisotzki$^{11}$ .\\
$^1$Department of Physics and Astronomy, University of British Columbia, 6224
Agricultural Road, Vancouver, V6T 1Z1, Canada.\\
$^2$Max-Planck-Institut f\"{u}r Astronomie, K\"{o}nigstuhl, D-69117,
Heidelberg, Germany.\\
$^3$Astronomisches Rechen-Institut, M\"onchhofstr.\ 12-14, D-69120 Heidelberg, Germany.\\
$^4$University of Texas, McDonald Observatory, Fort Davis, TX 79734, USA.\\
$^5$Department of Astronomy, University of Massachusetts, 710 North Pleasant
    Street, Amherst, MA 01003, USA.\\
$^6$Space Telescope Science Institute, 3700 San Martin Drive, Baltimore, MD
    21218, USA.\\
$^7$Department of Astrophysics, Denys Wilkinson Building, University of
    Oxford, Keble Road, Oxford, OX1 3RH, UK.\\
$^8$Johns Hopkins University, Baltimore, MD 21218, USA\\
$^9$Department of Astronomy, University of Texas at Austin, 1 University
    Station, C1400 Austin, TX 78712-0259, USA.\\
$^{10}$Centro Hispano Aleman de Calar Alto, C/Jesus Durban Remon 2-2, E-04004
    Almeria, Spain.\\
$^{11}$Astrophysikalisches Insitut Potsdam, An der Sternwarte 16, 14482 Potsdam,
    Germany.\\
}

\newcommand{\be}{\begin{equation}}  \newcommand{\ee}{\end{equation}}
\newcommand{\bes}{\begin{equation*}}  \newcommand{\ees}{\end{equation*}}
  \newcommand{\ba}{\begin{eqnarray}}
\newcommand{\ea}{\end{eqnarray}}

\newcommand{\rag}{\rangle}
\newcommand{\lag}{\langle}
\def\gs{\mathrel{\raise1.16pt\hbox{$>$}\kern-7.0pt %
\lower3.06pt\hbox{{$\scriptstyle \sim$}}}}         %
\def\ls{\mathrel{\raise1.16pt\hbox{$<$}\kern-7.0pt %
\lower3.06pt\hbox{{$\scriptstyle \sim$}}}}         %

\begin{document}

\pagerange{\pageref{firstpage}--\pageref{lastpage}} \pubyear{2005}
\maketitle
\label{firstpage}

\begin{abstract}

We present constraints on the evolution of the virial to stellar mass ratio 
of galaxies with high stellar masses in the redshift range $0.2<z<0.8$, by 
comparing weak lensing measurements of virial mass 
$M_{\rm vir}$ to estimates of stellar mass
$M_{\rm star}$ from COMBO-17.
For a complete sample of galaxies with 
$\log ( M_{\rm star} / M_\odot) > 10.5$,
where the majority show an early-type morphology, we find that the virial
mass to stellar mass ratio is given by
$ M_{\rm vir}/M_{\rm star} = 53^{+13}_{-16} $.  
Assuming a baryon fraction from the concordance cosmology, 
this corresponds to a stellar fraction of baryons in 
massive galaxies of $\Omega_b^*/\Omega_b = 0.10\pm 0.03$.
Analysing the galaxy sample in different
redshift slices, we find little or no evolution in the 
virial to stellar mass ratio, and place an upper limit of $\sim 2.5$ 
on the growth of massive galaxies through the 
conversion of gas into stars from $z=0.8$ to the present day.
\end{abstract}

\renewcommand{\arraystretch}{1.3}

\section{Introduction}

In the standard cold dark matter paradigm, luminous galaxies reside in
massive dark matter halos.  The large-scale clustering and 
formation of these luminous galaxies is driven
by the clustering and formation histories of their dark matter halos.
In contrast, the visible properties of such galaxies are a strong
function of the complex physics of baryonic matter,
in particular the laws governing stellar and AGN feedback, gas cooling and 
star formation.  The luminosities and colors of galaxies
are a strong function of the efficiencies of these processes, and 
our inability to robustly model them is a critical limitation of
current models of galaxy formation and evolution 
\citep[as discussed, for example, by][]{Benson03}.  
A key way to constrain the efficiency
of such processes is to measure empirically the ratio of stellar mass to
the total virial mass of the halo and galaxy system, for certain subsets
of the galaxy population.  Assuming a baryon to dark matter fraction 
from the concordant cosmology, it is then possible to
determine empirically the fraction of baryons which manage to cool 
and condense into stars in halos of a given mass.  Such measurements of 
the mass out to $\ge 100$\,kpc scales not only 
constrain the efficiency of complex `gastrophysical' processes, 
but they also empirically calibrate and test
halo occupation distribution analyses of galaxy properties 
\citep[see for example][]{Yang05}.

Satellite galaxies \citep{Zaritsky,Prada} and
weak gravitational lensing \citep[see the review by][]{Bible}, 
are currently the most promising methods available to probe
virial masses of dark matter halos.  
In this letter we focus on weak gravitational lensing,
the weak distortion of distant sources by
the gravitational field of foreground structure, 
as it provides a unique way to
probe matter on all scales, at all redshifts, 
irrespective of its dynamical state or nature.  Termed `galaxy-galaxy
lensing', the weak tangential shearing of background galaxies relative to
foreground galaxies can therefore
be used to probe the distribution of dark matter in galaxies, as
first demonstrated by \citet{BBS}. 

In this letter we combine information from two surveys of the Chandra 
Deep Field South; the
ground-based COMBO-17 survey \citep{Wolf03} and the HST GEMS
survey \citep{RixGEMS}, in order to constrain the 
virial to stellar mass ratio
in stellar massive galaxies out to $z \sim 0.8$.  
The 17-filter photometry of COMBO-17
provides accurate photometric redshifts with errors $\sigma_z \sim
0.02 (1+z)$.  COMBO-17 additionally provides stellar mass
estimates $M_{\rm star}$ by fitting a parameterised star formation history to the low
resolution 17-band galaxy spectra \citep{Borch}, based on a
\citet{Kroupa} initial mass function (Kroupa IMF).  The HST resolution 
of the GEMS imaging allows for high signal-to-noise measurements 
of the `galaxy-galaxy' lensing signal which can provide an ensemble
estimate of the virial mass for (sub-)sets of galaxies.  
A detailed description of the galaxy-galaxy
lensing analysis of GEMS 
and the resulting constraints on different dark matter halo
profiles for different galaxy types 
will be presented in a forthcoming paper (Heymans et al. in
prep).  For the purposes of this letter however we focus on a complete sample
of 626 lens galaxies with high stellar masses.
We measure the virial to stellar mass ratio as a function of galaxy 
redshift to $z \sim 0.8$, presenting the first redshift dependent
galaxy-galaxy lensing
analysis to be performed consistently over such a large redshift range.

This letter is organised as follows.  In Section~\ref{sec:ggtheory} we describe
the theory that underpins our measurement of virial masses and review the
particulars of our analysis.  We refer the reader
to section 3 of \citet{HymzGEMS} hereafter H05, for a lensing specific 
description of the GEMS
observations and the data reduction.  In Section~\ref{sec:results} we present
constraints on the virial radius and the 
virial to stellar mass ratio for our full
stellar massive galaxy sample and convert this into
a stellar fraction of baryons.   We compare our results to the low-redshift
results of \citet{HHbf05}, hereafter HHYLG, and \citet{RM06}, 
hereafter M06, and investigate the dependence of
the virial to stellar mass ratio on galaxy redshift, setting constraints 
on the growth of massive galaxies through the conversion of gas into stars 
from $z \sim 0.8$ to the present day.
We discuss the implications of these results and conclude in 
section~\ref{sec:conc}. 
Throughout this letter 
we assume a $\Lambda{\rm CDM}$ cosmology with $\Omega_m
= 0.3$, $\Omega_\Lambda= 0.7$ and $H_0 = 100 \, h {\rm km \, s}^{-1}\, {\rm
  Mpc}^{-1}$.    For our virial to stellar mass ratio measurement
we set $h=0.7$.

\section{Measuring virial masses with galaxy-galaxy lensing}
\label{sec:ggtheory}

The mass of a foreground `lens' galaxy will weakly shear the images of
background or `source' galaxies.  The extent of this tangential shearing 
as a function of lens and source position
can be directly derived from the density distribution of the lens galaxy.  
In this letter we adopt the NFW profile as a model for dark matter halos, where
the density distribution of a halo at redshift $z$ is
given by 
\be
\rho(r) = \frac{\delta_c \, \rho_c(z)}{(r/r_s)(1 + r/r_s)^2}\,.
\ee
$\delta_c$ is the characteristic density, $r_s$ is the scale radius and
$\rho_c(z)$ is the critical density\footnote{Note that
  within the galaxy-galaxy lensing 
literature $\rho_c$ is often defined differently. 
  In this letter we follow the
  definition given in the Appendix of \citet{NFW97}.} given by $3H(z)^2/
8\pi G$.  
The virial radius $r_{\triangle}$ is defined as the radius where the mass
density of the halo is equal to $\triangle \times \rho_c$.  \citet{Eke96} show that 
for flat $\Lambda$ cosmology, assuming a spherical collapse model,
\be 
\triangle(\Omega_m,\Lambda) = 178\, \Omega_m^{0.45} \,.
\label{eqn:triangle}
\ee
For the cosmology adopted here $\triangle \approx 100$
at $z=0$, and $\triangle \approx 150$ at $z=0.8$.
The corresponding virial 
mass\footnote{This
  definition of virial mass is directly equivalent to that used by HHYLG as 
 $\triangle(z) \rho_c(z)= \Delta_{\rm vir}(z) \rho_{\rm bg}(z)$ where
  $\Delta_{\rm vir}$ and $\rho_{\rm bg}$ are given after equation
  9 of HHYLG.} $M_{\triangle}$ is given by
\be
M_{\triangle} = \triangle \rho_c \frac{4\pi}{3} r_{\triangle}^3 \,.
\label{eqn:Mtriangle}
\ee
Defining the concentration parameter as $c = r_{\triangle} / r_s$, the
characteristic halo density $\delta_c$ is then given by
\be
\delta_c = \frac{\triangle}{3} \frac{c^3}{\ln(1+c) - c/(1+c)} \,.
\ee
Note that for a given CDM cosmology, halo mass $M_{\rm vir}$ 
and concentration $c$, or scale radius $r_s$, 
are related \citep{NFW97,Bullock01,Eke01}, where the
dependence is calculated through fits to numerical simulations.   

The expression for the weak lensing 
shear $\gamma$ induced by an NFW dark matter halo is given in \citet{WB00}.

\subsection{Lens galaxy sample selection}
\label{sec:sample}
To measure accurate virial masses it is important to have accurate
photometric redshifts.  To compare virial mass to stellar mass we also
require good stellar mass estimates.  
We therefore limit ourselves to a sample of lens 
galaxies with a measured COMBO-17
photometric redshift (this implies a magnitude limit of $R<24$)
between $0.2 < z < 0.8$.  For the purposes of this letter we wished
to create a complete sample of galaxies in order to be able to study
the virial to stellar mass ratio as a function of redshift without the
inclusion of potentially biasing selection
effects.  This requirement imposed a high stellar
mass cut where $\log ( M_{\rm star} / M_\odot) > 10.5$, 
above which it has been shown that the COMBO-17 catalogue is complete in both
redshift estimates and flux \citep{C17,Borch}.  The S\'ersic 
fits of these galaxies \citep[see][]{Barden,McIntosh} 
indicate that the majority of the lens sample ($ \sim 75\%$)
are early-type galaxies with $n>2.5$.  
Note that including an $n>2.5$ 
S\'ersic selection criteria such that our lens sample contains only 
early-type galaxies does not change our results except for
decreasing the lensing signal-to-noise.
The galaxy sample has an average photometric redshift accuracy of
$\delta_z \sim 0.03$ and a mean stellar mass of 
$\langle M_{\rm star} \rangle = 7.2 \times 10^{10} M_\odot $. 
The galaxies are very luminous with a mean SDSS $r$-band luminosity 
$\langle L \rangle = 2.4 \, L_*$, 
where $L_* = 10^{10} h^{-2} L_\odot$.  They also have a 
low dynamic range of galaxy luminosity with $\sigma(\log L/L_*) = 0.2$. 

Source galaxies with which to measure the galaxy-galaxy lensing signal
are selected from the GEMS data as 
described in section 4.3 of H05.  This yields a number density of 65 source 
galaxies per square arcmin. 

The lens redshift limits were chosen to optimise the measured signal.
Galaxies at 
higher redshifts typically 
have poorer photometric redshift accuracy and they also
provide rather weak constraints owing to the reduced number of
background sources.  The lower redshift limit results from the fact that we
exclude all sources for which lenses might lie 
outwith the GEMS area. For a given maximum impact parameter
$r_{\rm max}$, this corresponds to a minimum
angular separation
$\theta(z_{l\, {\rm min}})$ from the edge of the field.  We therefore include
the lower redshift limit to reduce $\theta(z_{l\, {\rm min}})$ thus including
as many sources as possible.    

In this letter we will make two assumptions.  As our lens sample spans a 
low dynamic range in luminosities we will assume that the halo 
virial radius $r_\triangle$ 
of our lens sample is constant.  M06 find that stellar mass 
is a good galaxy-type-dependent proxy for halo mass.
We will therefore also assume that,
on average, the halo virial
mass of our galaxy sample scales linearly with galaxy stellar mass by 
defining the following scaling relation,
\be
M_{\rm vir} = k_{\rm NFW} M_{\rm star} \,.
\ee
Using the mass/concentration relations from \citet{NFW97}, 
the halo profile of a galaxy in our sample 
can then be uniquely described by an NFW model with one free parameter; 
either $r_{\triangle}$ or $k_{\rm NFW}$.  
  
\subsection{Analysis Method}
\label{sec:method}

We use the method of \citet{KSB}, \citet{LK97} and \citet{HFKS98} (KSB+) 
to obtain an estimate of the observed ellipticity $\epsilon^{\rm obs}$ 
of each source galaxy in the GEMS survey.  In the weak shear limit, 
where $\gamma \ll 1$, 
the observed ellipticity $\epsilon^{\rm obs} \approx \epsilon^{\rm s} +  
\gamma$, where $\epsilon^{\rm s}$ is the intrinsic 
source galaxy ellipticity.   The
application of the method to the GEMS data set is described in detail in
H05 along with the results of 
several diagnostic tests which confirm the
success of this method at removing PSF distortions.  The
levels of shear 
calibration bias expected with the H05 implementation of KSB+ has
been shown to be $\sim 3\%$ \citep{HymzSTEP}
which is well within the statistical uncertainties of this analysis.  

In order to interpret the measured average tangential shear
in the context of NFW halo lensing, redshift information for both the
lens and source galaxies is required \citep[see for example][]{MK05}.
The COMBO-17 survey provides accurate photometric redshifts for the foreground 
lens sample of galaxies with $R<24$ \citep{C17}.  
For the fainter source galaxy sample only the
redshift probability distribution is known
(see section 6 of H05).  
The maximum-likelihood method of \citet{SchRix} was designed to take advantage
of such a data set by analysing the galaxy-galaxy lensed shear
statistically, and it is this method that we use in the following 
analysis and describe below.   

For a model galaxy density profile, in the case where all galaxy 
redshifts are known, 
the weak shear $\gamma$ experienced by 
each source galaxy can be predicted by 
summing up the shear contributions from all
the foreground lens galaxies, taking into account the
multiple weak deflections that a source may experience \citep{Bmultdef06}.    
Note that when a sample of lenses are selected 
from the foreground galaxy population, the other foreground galaxies 
act as a source of noise without introducing bias \citep{MK06}.
In this analysis the redshifts of the source galaxies are unknown, and
we therefore assign those galaxies a magnitude-dependent redshift  
probability distribution $p(z,{\rm mag})$ (equation 15 H05) and calculate the
expectation value of the shear $\lag \gamma \rag$.  This is done through Monte
Carlo integration, drawing a source galaxy redshift estimate $z_s^\nu$
from the distribution $p(z,{\rm mag})$, $\nu = 1..N_{\rm MC}$ times, where
$N_{\rm MC} = 100$ in this analysis.  For each
$z_s^\nu$ estimate the induced galaxy-galaxy lensing shear $\gamma^\nu$ 
is calculated with the resulting mean shear given by
\be
\lag \gamma \rag = \frac{1}{N_{\rm MC}} \displaystyle\sum_{\nu=1}^{N_{\rm MC}}
\gamma^\nu \,.
\ee
Note that in practice we only consider the shear from 
those lenses where the angular separation from the source $\theta <r_{\rm
  max}/D_l$ with $r_{\rm max} = 300 h^{-1}{\rm kpc}$. The intrinsic 
source galaxy ellipticity $\epsilon^{\rm s}$ is then calculated,
$\epsilon^{\rm s} \approx
\epsilon^{\rm obs} - \gamma$. The distribution of each component of the 
observed galaxy ellipticity
is well described, for the GEMS survey,
by a Gaussian of width $\sigma_\epsilon = 0.31$.
As the induced shear $\gamma$ is weak, the probability for observing an
intrinsic ellipticity of $\epsilon^{\rm s}$ is then given by
\be
P(\epsilon^{\rm s}) = \frac{1}{2\pi\sigma_\epsilon^2} \exp \left[
  \frac{-|\epsilon^{\rm s}|^2}{2 \sigma_\epsilon^2} \right] \,\, .
\ee
The best-fit dark matter halo parameters are determined
by maximising the likelihood $L = \Pi \left[P(\epsilon^{\rm s})_i\right]$ 
where the product extends over all source galaxies $i$.

\section{Results}
\label{sec:results}
\begin{figure}
\begin{center}
\epsfig{file=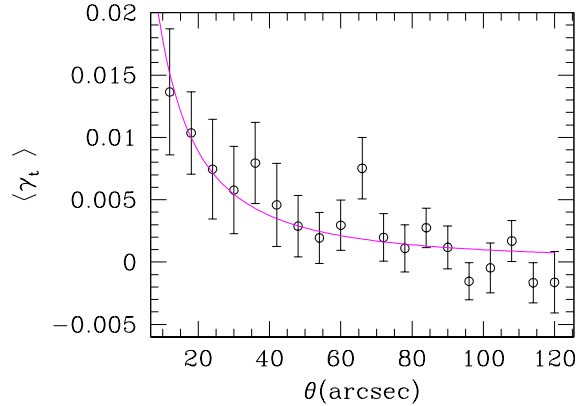,width=7.5cm,angle=0,clip=}
\caption{The average tangential 
alignment $\langle \gamma_t \rangle$ of source galaxies with respect to the 
lens sample as a function of angular separation on the sky $\theta$.  
The solid curve
shows the theoretical lensing signal from 
an NFW halo at $z=0.65$, the median redshift of the lens sample,
with the maximum likelihood method constrained 
virial radius $r_\triangle = 204 h^{-1} {\rm kpc}$.}
\label{fig:et}
\end{center}
\end{figure}

In this section we present our constraints on the virial radius and the 
virial to stellar mass ratio of the stellar massive galaxies in our lens 
sample.
Assuming that the chosen subset of lens galaxies have a constant virial radius
we find $r_\triangle = 204^{+18}_{-22} h^{-1} {\rm kpc}$. 
This corresponds to a virial mass of 
$M_{\rm vir} = 29.8^{+7.9}_{-9.6} \, 10^{11} h^{-1} {\rm M}_\odot$ 
for a galaxy at $z \sim 0.65$, the median redshift of the lens sample.
Note that the corresponding 
virial mass is dependent on the redshift of the halo
because of the redshift dependence of 
the critical density $\rho_c$ and $\triangle$ (equation~\ref{eqn:Mtriangle}).
For our mean galaxy luminosity we find an average 
mass-to-light ratio for the stellar massive galaxies of 
$M_{\rm vir}/L = 123 \pm 36 \,h {\rm M}_\odot/{\rm L}_\odot$.
To illustrate this result, obtained using the maximum likelihood method 
described in section~\ref{sec:method}, figure~\ref{fig:et} shows the 
average tangential 
alignment $\langle \gamma_t \rangle$ of source galaxies with respect to the 
lens sample as a function of angular separation on the sky.  The measured 
signal $\langle \gamma_t \rangle$ is well fit by the expected signal around
an NFW halo at the median redshift of the lens sample 
with the maximum likelihood method constrained 
virial radius of $r_\triangle = 204 h^{-1} {\rm kpc}$.  

Our results are in good agreement 
with the results of HHYLG who find 
$M_{\rm vir} = 23.3 ^{+5.6}_{-5.1}\times 10^{11} h^{-1} {\rm M}_\odot$ 
for their luminous
galaxy sample at $z \sim 0.32$, where $\langle L \rangle = 2.49 L_*$.
Scaling our results to $z=0.32$ we find 
$M_{\rm vir}= 17.8^{+4.7}_{-5.7}\times 10^{11} h^{-1} {\rm M}_\odot$.
These results are also in good agreement with the results of M06 who find
a virial mass for central galaxies (as opposed to satellite galaxies) of
$M_{\rm cent} = 14.1^{+5.6}_{-5.3} \, 10^{11} h^{-1} {\rm M}_\odot$,  
for a sample of early-type galaxies with mean stellar mass,
$\langle M_{\rm star}\rangle = 5.8 \times 10^{10} M_\odot$ and 
mean luminosity $\langle L \rangle \sim 1.3 L_*$.
This galaxy sample has a mean redshift $z \sim 0.11$, and scaling our 
results to this redshift we find a consistent virial mass of
$M_{\rm vir} = 12.5^{+3.3}_{-4.0} 10^{11} h^{-1} {\rm M}_\odot$
for our sample of stellar massive galaxies with mean stellar mass,
$\langle M_{\rm star}\rangle = 7.2 \times 10^{10} M_\odot$ and 
mean luminosity $\langle L \rangle \sim 2.4 L_*$.   

\begin{figure}
\begin{center}
\epsfig{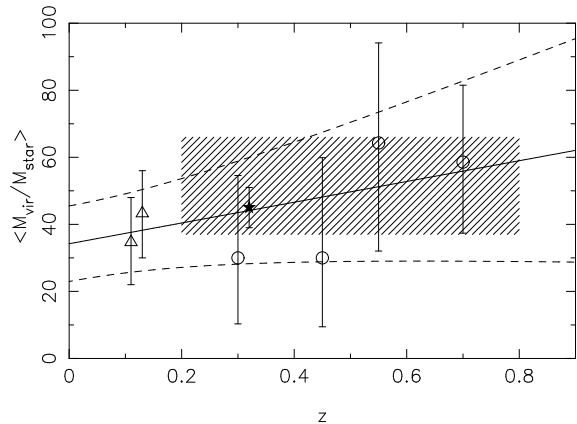}
\caption{The virial mass to stellar mass ratio as a function of galaxy
redshift (circles).  The result at $z=0.32$ is from HHYLG (star).  
The results at $z=0.11$ and $z=0.13$ are 
from M06 (triangle) measured from different stellar mass selected 
samples that best match our lens sample.  
The errors are $1\sigma$ in all cases, with the exception M06 who
quote $3 \sigma$ errors.  
The hatched region is the $1\sigma$ confidence regions for the 
virial mass to stellar mass ratio measured from GEMS
over the full redshift range $0.2 < z < 0.8$.  
The solid line indicates the best linear fit to 
$M_{\rm vir}/M_{\rm star}(z)$ and the dashed lines denote the $1\sigma$ 
errors to this fit. }
\label{fig:kvsz}
\end{center}
\end{figure}

In Section~\ref{sec:sample} we introduced a stellar mass scaling to relate
halo mass to stellar mass, which we
choose to be linear\footnote{Note that we also tested non-linear stellar mass 
scaling, but found for
our complete sample of $\log(M_{\rm star} / M_\odot)>10.5$ 
galaxies, any non-linear component
was consistent with zero within the noise}. 
For the full lens sample we measure a best fit value for the NFW virial mass to
stellar mass ratio as $k_{\rm NFW}=53^{+13}_{-16}$ with $1\sigma$
errors.  Note that $k_{\rm NFW}=1$ is ruled out with 99.99\% confidence.
Our result is in good agreement with the results of
HHYLG and M06, as shown in Figure~\ref{fig:kvsz}.
HHYLG find $k_{\rm NFW} = 45 \pm 6$ for a scaled Salpeter IMF which
is similar to the Kroupa IMF used in this 
analysis\footnote{Note that the Kroupa IMF
and scaled Salpeter IMF are roughly half the mass of a Salpeter IMF 
and hence our results would scale to $k_{\rm NFW}^S \sim 27$ in comparison
to the HHYLG Salpeter IMF measurement of $k_{\rm NFW}^S = 26 \pm 4$.}.
M06 find $k_{\rm NFW}= 35 \pm 13$ using a Kroupa IMF,
for the sample of early-type galaxies whose average stellar mass best
matches the average stellar mass of our sample, as discussed above. 
For a higher stellar mass selected sample with 
$\langle M_{\rm star}\rangle = 11.2 \times 10^{10} M_\odot$, M06 find
$k_{\rm NFW}= 43 \pm 13$.  As $\sim 20\%$ of our sample would fall into this
mass bin, we also show this result in Figure~\ref{fig:kvsz}

Following the analysis of HHYLG and M06 
we convert our virial to stellar
mass ratio into a stellar baryon fraction $f_*$ 
by assuming that the baryon to dark
matter ratio in massive galaxies is given by the global value
$\Omega_b / \Omega_m = 0.176 \pm 0.013$ \citep{WMAP06} such that
\be
f_* = \frac{M_{\rm star}}{M_{\rm vir}} \frac{\Omega_m}{\Omega_b}
\sim \Omega_b^*/\Omega_b \,\, .
\ee
The measured virial to stellar mass ratio 
then gives us a stellar baryon fraction of
$f_* = \Omega_b^*/\Omega_b = 0.10 \pm 0.03$, 
suggesting that galaxies with high stellar masses 
are rather inefficient at converting 
baryons into stellar mass: only 10\% of the baryons are contained in stars.
This result is in good agreement with the cosmic baryon budget 
of \citet{Fukugita} who estimate the 
the amount of baryons in different states, finding the fraction of 
baryons in spheroids to be $f_* \sim 0.12$.  

\subsection{Redshift dependence}

Splitting our galaxy sample into four redshift slices
we show, in Figure~\ref{fig:kvsz}, that the virial to stellar mass
ratio remains fairly constant as a function of galaxy redshift.  For a
linear fit to these results, together with the estimates of HHYLG and M06,
we can constrain $k_{\rm NFW}(z) = (34 \pm 11) + (31 \pm 35) z$. 
Comparing the virial to stellar mass ratio at $z=0$ and $z=0.8$, 
we find a $1\sigma$ confidence region where
$0.8<(k[z=0.8] / k[z=0])<2.6$. The data therefore indicate that
the virial to stellar mass ratio has evolved by a factor $\lesssim 2.5$ 
since z=0.8.

\section{Discussion and Conclusion}
\label{sec:conc}

In this letter, we use a weak gravitational lensing study of the 
HST GEMS survey, in combination with the 17-colour photometry 
of the COMBO-17 survey, to determine the 
ratio between virial and stellar mass in a complete sample of massive
galaxies out to $z \sim 0.8$.  We find a mean virial to stellar mass ratio
$M_{\rm vir}/M_{\rm star}=53^{+13}_{-16}$ with 
little or no redshift evolution in
this stellar mass fraction.  

This result provides a high redshift extension
to the analysis of \citet{HHbf05} (HHYLG) and \citet{RM06} (M06).
The excellent agreement found between the results of this 
analysis, HHYLG and M06 
is important to comment upon 
as the analyses were performed using very different
techniques and came from very different surveys: HHYLG use the Red 
sequence Cluster Survey, a medium-deep
$45.5$ square degree ground-based 
survey in $BVRz$, M06 use the shallow Sloan Digital all Sky Survey which 
is ground-based in $ugriz$. This analysis uses GEMS, 
a deep $0.25$ square degree space-based survey
with accompanying 17-band COMBO-17 data.

In this analysis we use stellar masses calculated from the
best-fitting models of star formation history to the
low resolution 17-band galaxy spectra.  M06 use a similar technique 
\citep{Kauffmann03}.
HHYLG use the simple
relationship of \citet{BelldeJong} that relates stellar mass to $B-V$
colour.  This relationship has a systematic uncertainty that is often 
estimated to be $\sim 30\%$ \citep{Bell03}.  Stellar mass uncertainty is 
indeed hard to avoid with all methods, however, 
owing to the unknown frequency of 
bursts of star formation and uncertainty in dust models \citep{Borch}.
The fact all three results are in such close agreement, however,
demonstrates that the simple models of HHYLG are consistent with the more 
complex star formation history models of COMBO-17 and M06, which bodes well for
similar studies in future broad-band surveys.

We have applied the \citet{SchRix} maximum likelihood
method to set constraints on the virial radius and the virial to stellar 
mass ratio using an NFW dark matter profile.  This method 
takes into account the uncertainty in the redshift of both the
source and lens galaxies, using a Monte Carlo technique 
and includes the effects of the multiple deflection of
light by successive foreground lens galaxies.  
In contrast, HHYLG measure the
tangential shear around foreground galaxies for different luminosity
bins and fit this result with the shear expected from an NFW profile.  
The HHYLG galaxies are selected to be isolated such that 
the lensed background galaxies experience only the gravitational
potential of the isolated galaxy.  
Their larger photometric redshift errors are more likely to
scatter faint galaxies into higher luminosity bins than scatter bright
galaxies into lower luminosity bins.  This then leads to a bias in the
measurement of the mass at
fixed luminosity to a lower value.  HHYLG therefore need to 
apply a correction factor 
to the virial masses which they determine from an analysis of mock
catalogues.  This bias is not a problem for our analysis owing to the
higher accuracy of photometric redshifts and the use of the
\citet{SchRix} method.  Our observational 
systematic errors are therefore minimal, 
and we are limited by random errors from smaller number statistics.
The third analysis method of M06 uses the 
halo model to extract information from the 
galaxy-galaxy lensing signal and separate contributions from central and 
satellite galaxies \citep{Mandelbaumgg}.
The fact that these three very different methods produce consistent 
results serves as a good indication that the three results are robust.

This analysis is the
first time a comparison of the properties of luminous galaxies 
and their parent dark matter halos has been 
performed consistently across a substantial span of
cosmic time.  The galaxy sample has been selected to be complete out to
$z \sim 0.8$ and we are therefore not subject to redshift-dependent
selection biases.   
We have shown that for this galaxy sample, the majority of which show 
early-type morphologies, the
virial to stellar mass ratio as a function of galaxy redshift
remains fairly constant with $M_{vir}/M_{\rm star} \sim 50$: 
specifically, we place a $1 \sigma$ upper 
limit on the growth of the stellar to virial mass ratio
since $z=0.8$, of a factor $< 2.6$.  Because we probe the ratio of stellar
to virial mass, we cannot probe merger-driven growth of galaxies as stellar
masses and virial masses grow in lock-step.  Instead, our constraint 
effectively limits the conversion of gas into stars in the massive 
galaxy population, i.e. star formation, 
as a function of redshift. We therefore can conclude that the 
growth of stellar mass due to star formation is limited, 
to a factor of $\lesssim 2.5$ from $z=0.8$ to the present day. 

\section{Acknowledgments}
We thank Julio Navarro for making the NFW 
charden and ENS codes publicly available.  We also wish to thank 
Martina Kleinheinrich for useful discussions and the referee for
his/her helpful comments.
CH is supported by a CITA National fellowship, and with
HWR acknowledges financial support from GIF. EB is supported
by the DFG's Emmy Noether Programme, CW by PPARC,
CYP by STScI, SJ and DHM by NASA under LTSA Grant NAG5-13063(SJ) and
NAG5-13102(DHM). Support for GEMS was provided by NASA through 
GO-9500 from STScI operated by AURA under NAS5-26555.

\bibliographystyle{mn2e}
\bibliography{ceh_2005}
\label{lastpage}

\end{document}